\documentclass{article}

\usepackage[nonatbib,preprint]{neurips_2025}

\usepackage[utf8]{inputenc} 
\usepackage[T1]{fontenc}    
\usepackage{hyperref}       
\usepackage{url}            
\usepackage{booktabs}       
\usepackage{amsfonts}       
\usepackage{nicefrac}       
\usepackage{microtype}      
\usepackage{xcolor}         
\usepackage{graphicx}
\usepackage{amsmath}
\usepackage{bm}
\usepackage{multirow}       
\usepackage{comment}

\title{INN-FF: A Scalable and Efficient Machine Learning Potential for Molecular Dynamics}

\author{%
  Taskin Mehereen \\
  Department of Mechanical Engineering\\
  Bangladesh University of Engineering and Technology\\
  Dhaka, Bangladesh \\
  \texttt{1910124@me.buet.ac.bd} \\
  \And
  Sourav Saha \thanks{Corresponding author} \\
  Kevin T. Crofton Department of Aerospace and Ocean Engineering \\
  Virginia Polytechnic Institute and State University\\
  Blacksburg, VA 24060 \\
  \texttt{souravsaha@vt.edu} \\
  \And
  Intesar Jawad Jaigirdar \\
  Department of Computer Science and Engineering \\
  University at Buffalo \\
  Buffalo, NY 14260 \\
  \texttt{intesarj@buffalo.edu} \\
  \And
  Chanwook Park \\
  Department of Mechanical Engineering \\
  Northwestern University \\
  Evanston, IL 60208\\
  \texttt{chanwookpark2024@u.northwestern.edu} \\
}

\begin{document}

\maketitle

\begin{abstract}
  The ability to accurately model interatomic interactions in large-scale systems is fundamental to understanding a wide range of physical and chemical phenomena, from drug-protein binding to the behavior of next-generation materials. While machine learning interatomic potentials (MLIPs) have made it possible to achieve ab initio-level accuracy at significantly reduced computational cost, they still require very large training datasets and incur substantial training time and expense. In this work, we propose the Interpolating Neural Network Force Field (INN-FF), a novel framework that merges interpolation theory and tensor decomposition with neural network architectures to efficiently construct molecular dynamics potentials from limited quantum mechanical data. Interpolating Neural Networks (INNs) achieve comparable or better accuracy than traditional multilayer perceptrons (MLPs) while requiring orders of magnitude fewer trainable parameters. On benchmark datasets such as liquid water and rMD17, INN-FF not only matches but often surpasses state-of-the-art accuracy by an order of magnitude, while achieving significantly lower error when trained on smaller datasets. These results suggest that INN-FF offers a promising path toward building efficient and scalable machine-learned force fields.
\end{abstract}

\section{Introduction}
\label{intro}
Accurate prediction of atomic interactions is fundamental for understanding physical and chemical phenomena, from drug-protein binding to the design of advanced materials \cite{huan2017}. While quantum mechanical methods such as density functional theory (DFT) offer the required accuracy, their prohibitive computational cost makes them impractical for studying large systems over extended time and length scales \cite{vennelakanti2021}. 

To overcome these limitations, molecular dynamics (MD) simulations have emerged as a practical tool. However, the accuracy and reliability of MD are fundamentally tied to the quality of the underlying force fields (FFs). A force field is a mathematical model that approximates the potential energy surface of a system, allowing the calculation of forces between atoms based on their positions. Traditional force fields are often highly specialized, requiring extensive parameter tuning and suffering from limited transferability, especially when modeling new materials or chemical environments.

Machine-learned force fields (MLFFs) have recently shown promise in bridging this gap, achieving quantum-level accuracy while enabling efficient simulations\cite{noe2020}.  Yet, these approaches remain heavily dependent on large volumes of high-quality training data, making them costly and sometimes impractical for less-explored material spaces.

In this work, we introduce the Interpolating Neural Network Force Field (INN-FF), a novel machine learning framework designed to overcome the data inefficiency bottleneck in existing MLFFs. By using interpolation theory and tensor decomposition within a neural network architecture, INN-FF achieves high accuracy with significantly fewer training samples and trainable parameters, offering a scalable and efficient path for molecular simulations without compromising fidelity.

\section{Background and related work}
\label{related_work}
\noindent To address the challenges of modeling atomic interactions efficiently, parametric force fields were developed, using simplified mathematical expressions to approximate interatomic forces. Traditional force fields, such as the Lennard-Jones potential, the Embedded Atom Method (EAM), and its modified form (MEAM), have been foundational in molecular and materials simulations. The Lennard-Jones potential is particularly effective for modeling van der Waals and nonbonded interactions in simple systems~\cite{Jones1924}, while EAM and MEAM incorporate many-body effects to better describe metallic bonding~\cite{Daw1984, Baskes1992}. For biomolecular and organic systems, force fields like GAFF, CHARMM, and OPLS-AA partition the total energy into bonded and nonbonded contributions, enabling large-scale simulations~\cite{Wang2023, Greener2024, Maginn2023}. However, their fixed functional forms and manually tuned parameters often fail to capture the complex interactions necessary for bond breaking, charge transfer, or long-range electronic effects. Moreover, parameterization is labor-intensive and limits transferability across diverse chemical environments~\cite{Chen2023}.

\noindent Recent advances in machine learning (ML) have begun to address these limitations, demonstrating the ability to reproduce quantum mechanical accuracy while significantly reducing computational costs~\cite{noe2020}. Machine-learned force fields (MLFFs) have opened new avenues in simulations, offering quantum-level fidelity across areas from protein folding to catalysis~\cite{biamonte2017}. MLFFs such as Gaussian Approximation Potentials (GAP) and Neural Network Potentials (NNP) leverage high-quality DFT datasets, and frameworks like SchNetPack, M3GNet, and AMP employ graph neural networks (GNNs) and message passing to model complex interatomic interactions~\cite{Zhang2018, Chen2022, Fu2023}. Recent innovations like MACE~\cite{Batatia2023} push this further, using higher-order equivariant message passing to improve both scalability and accuracy.

\noindent Graph-based models such as M3GNet generalize effectively across chemistries and structures by leveraging large datasets of relaxed structures~\cite{Chen2022}, while architectures like Kolmogorov-Arnold Networks (KANs) offer a mathematically principled, interpretable alternative to traditional GNNs through learnable edge activations~\cite{Liu2024}. Despite these advances, a persistent inefficiency remains: the evaluation of atomic forces, critical for molecular dynamics, often incurs significant overhead due to complex network architectures~\cite{karplus2005}.

\section{The INN architecture}
\label{architecture}
Interpolating Neural Networks (INNs) attain exceptional efficiency in terms of model accuracy given the number of trainable parameters through two main features: adaptable nonlinear interpolation functions rooted in locally supported interpolation theories inspired by finite element methods and tensor decomposition (TD)-based model order reduction. The former constructs a learnable 1D activation function similar to KANs. While KANs adopt splines as the basis, INNs use general interpolation techniques such as Lagrange polynomials, HiDeNN \cite{zhang2021hierarchical, saha2021hierarchical}, and C-HiDeNN \cite{lu2023convolution, park2023convolution}. TD methods, specifically CANDECOMP / PARAFAC (CP) \cite{harshman1970foundations} decomposition, then merge the outputs of 1D activation functions to construct a high-dimensional approximation space using only a single hidden layer architecture. While MLPs and KANs typically require deep network architectures for practical applications, the single-layer architecture of INN offers remarkable scalability for high-dimensional problems. Together, these components significantly strengthen the model’s ability to represent complex input-output systems with greater accuracy and reduced computational cost.

\subsection{One-dimensional INN}

Consider a univariate function $u(x)$ defined in an input domain $x\in[0,1]$. A linear finite element method approximates this function by 1) discretizing the domain with $J$ nodes ($J-1$ elements) where $x^{(j)}$ being the nodal coordinates, 2) building linear interpolation functions on each node $N^{(j)}(x)$, and 3) interpolating the function values measured at nodes $u^{(j)}=u(x^{(j)})$. This 1D approximation $u(x)\approx u^h(x)$ can be written as:

\begin{equation}
\label{eq:hidenn}
    u(x) \approx u^h (x)=\sum_{j=1}^{J} {N^{(j)}(x)u^{(j)}}={\bm{N}(x)\cdot\bm{u}},
\end{equation}
where $\bm{N}(x), \bm{u}\in\mathbb{R}^J$ are the vector forms of the interpolation functions and nodal values, respectively.

\begin{figure}[h]
    \centering
    \includegraphics[width=1.0\linewidth]{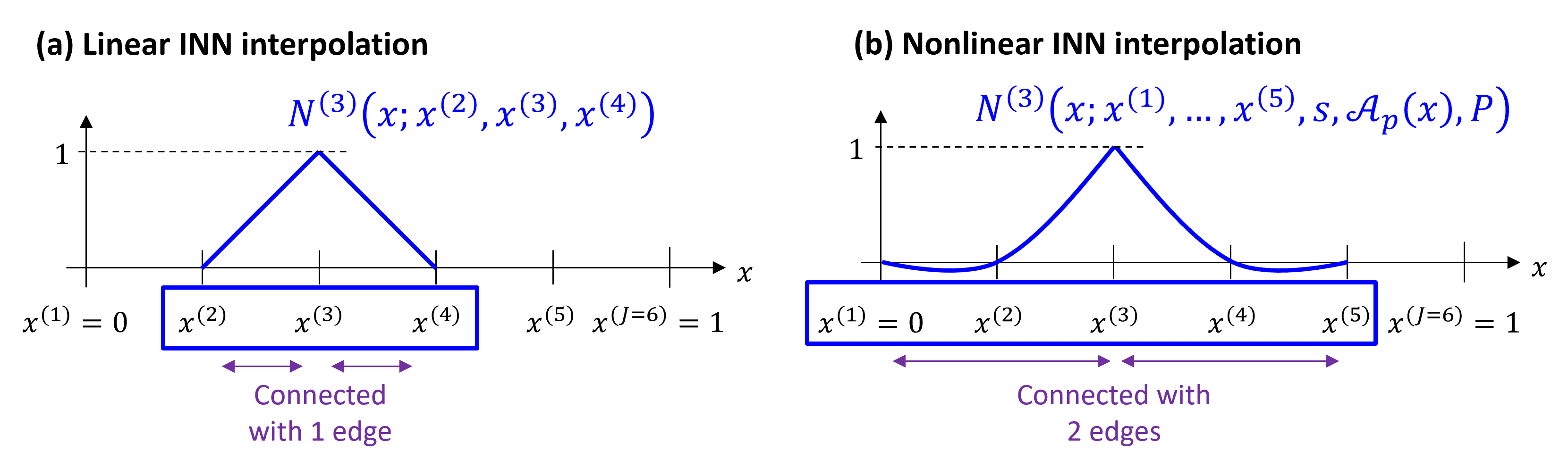}
    \caption{INN interpolation functions in 1D input space for linear (a) and nonlinear (b) cases. The input domain $x\in[0,1]$ is divided into $J=6$ nodes. Interpolation functions at node $j=3$ are visualized with the size of the support domain (i.e., patch size $s$). In (b), nonlinear basis functions $\mathcal{A}_p(x), p=1,\dots,P$ are also needed to construct interpolation functions.}
    \label{fig:INN_1D}
\end{figure}

As described in Figure \ref{fig:INN_1D}(a), the linear INN interpolation function at node $j=3$ is a compact-supported hat function. The ingredients for this function are the nodal coordinates: the center node $x^{(3)}$ and the one layer of neighboring nodes: $x^{(2)}, x^{(4)}$. 

The nonlinear INN interpolation function in Figure \ref{fig:INN_1D}(b) has a larger support domain whose size is characterized by patch size $s=2$ (i.e., two layers of neighboring nodes from the center node $x^{(3)}$). It has additional adaptable hyperparameters, INN basis functions $\mathcal{A}_p(x), p=1,\dots,P$, where the number of basis functions $P$ is limited by the patch size: $P\leq s$. INNs can adapt these hyperparameters to improve approximation accuracy without changing the total number of trainable parameters, which in this case is $J=6$. A detailed procedure for computing these functions is provided in Appendix A.1. Notice that the INN interpolation functions satisfy the following properties of an interpolation function: 1) compact-supported, 2) partition of unity, 3) Kronecker delta, 4) reproducing condition \cite{lu2023convolution}. 

\begin{figure}[h]
    \centering
    \includegraphics[width=1.0\linewidth]{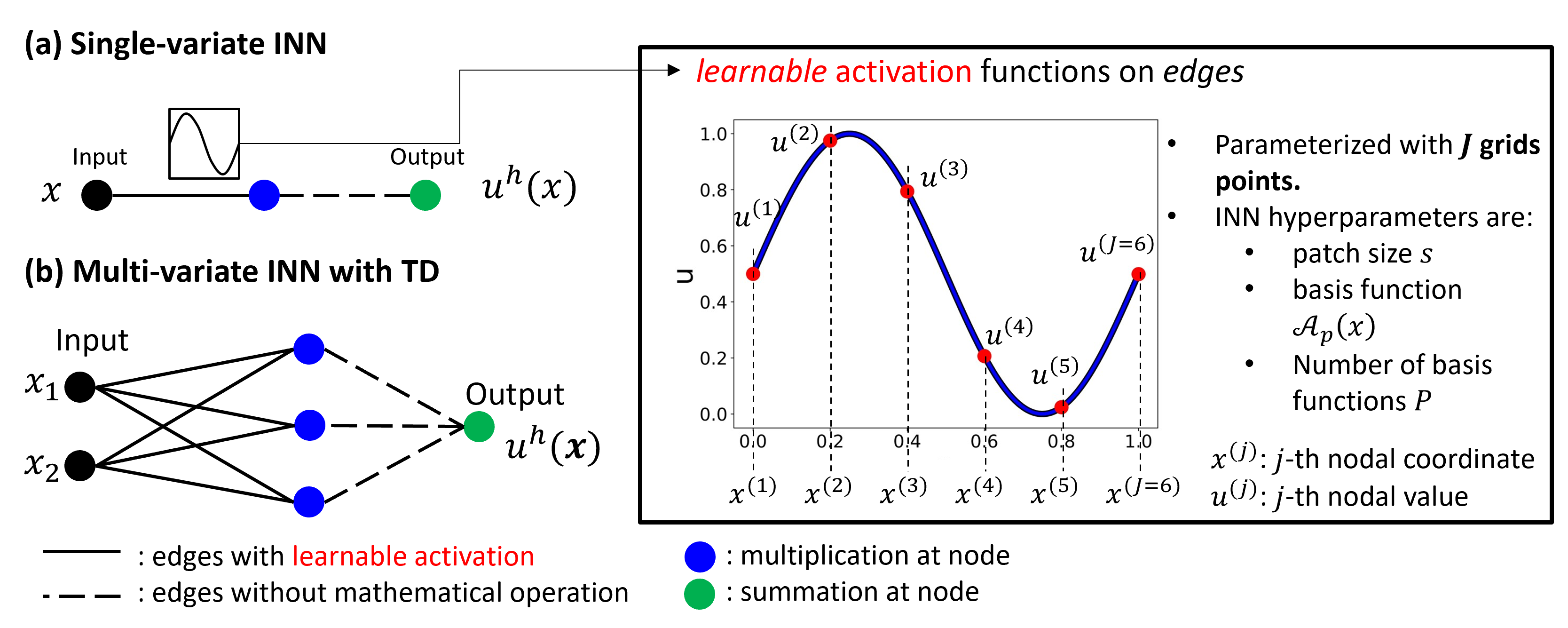}
    \caption{INN network architecture for 1D input variable (a) and 2D input variables (b). TD: Tensor Decomposition.}
    \label{fig:INN_TD}
\end{figure}

The corresponding network architecture is provided in Figure \ref{fig:INN_TD}(a) for the single-variate case. This one-input one-output system has one hidden layer with one hidden neuron. The edge from the input neuron to the hidden neuron is represented as a learnable activation function parameterized with $J$ trainable parameters (i.e., nodal values).

\subsection{INN approximation with TD}

For multivariate functions, INNs utilize CP tensor decomposition (TD) to ensure that the number of trainable parameters scales linearly with the number of input variables. Consider an $I$-dimensional function $u(x_1,x_2,\dots,x_I)\in\mathbb{R}$. An INN approximates it as a finite summation of products of univariate INNs on each input variable:

\begin{equation}
    u(x_1,\dots ,x_I)\approx u^h(x_1,\dots ,x_I)=\sum_{m=1}^{M}\prod_{i=1}^{I}\bm{N}_i(x_i)\cdot \bm{u}_i^{(m)},
    \label{td_eq}
\end{equation}
where $\bm{N}_i(x_i), \bm{u}_i^m\in\mathbb{R}^J$ are the 1D INN interpolation functions and the corresponding nodal values in the $i$-th input and for $m$-th TD mode. 

From Figure \ref{fig:INN_TD}(b), one can infer that the number of TD modes $M$ is revealed as the number of hidden neurons. The number of trainable parameters of an INN is thus $MIJ$, which scales linearly with the number of neurons. This is different from MLPs and KANs where the parameters scale quadratically with the number of neurons due to the deep network architecture \cite{Park2024}. This unique feature of INN enables exceptional scalability and efficiency compared to MLPs or KANs.


\section{Results}
\label{results}
\subsection{Scalability and data efficiency}
\label{scalable}
\begin{figure}[h]
    \centering
    \includegraphics[width=0.7\linewidth]{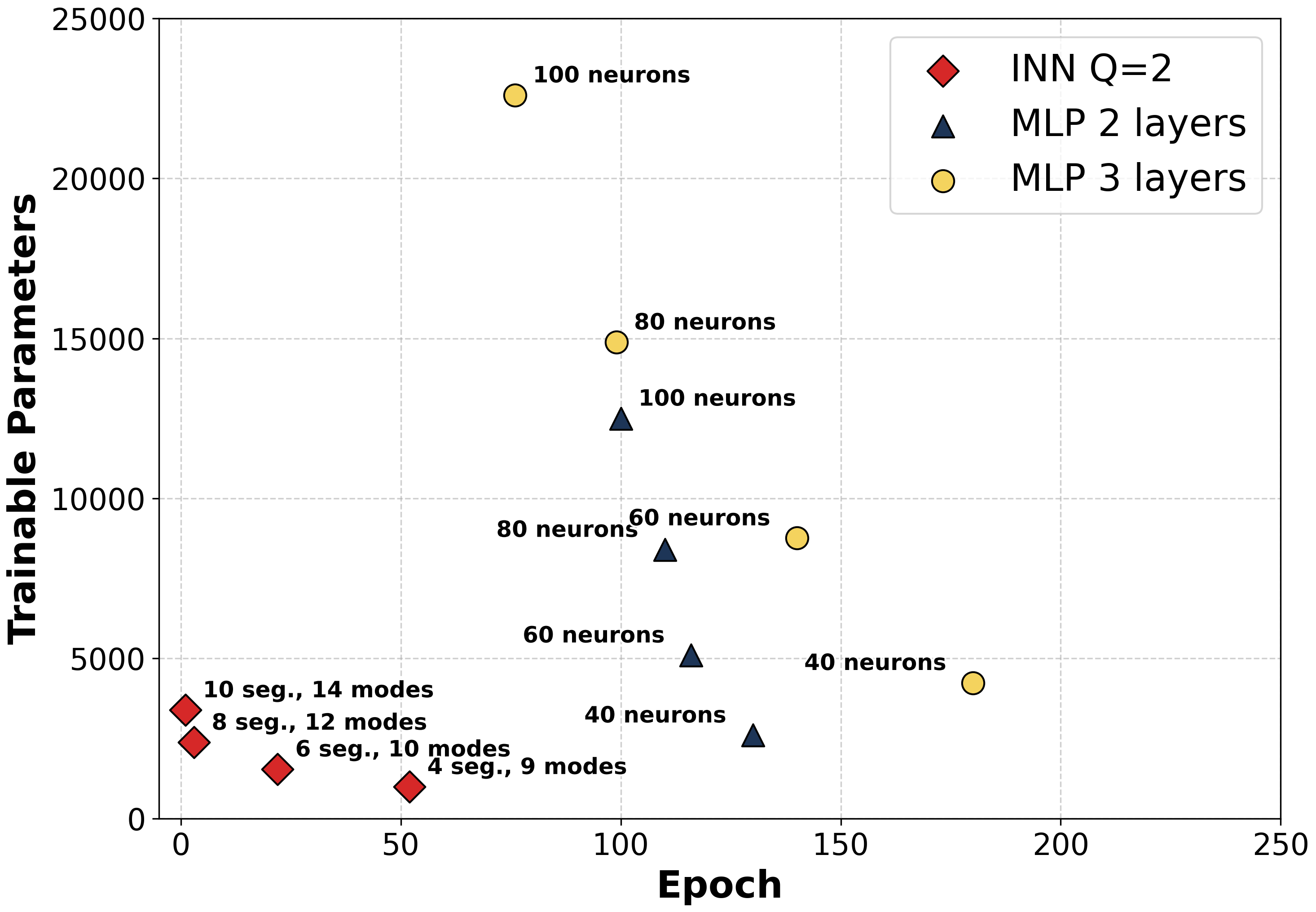}
    \caption{Trainable parameters vs. epochs. INNs trained on bulk water dataset \cite{Cheng2019} consistently achieve faster convergence with fewer parameters compared to traditional MLPs.}
    \label{fig:params_vs_epochs}
\end{figure}

\paragraph{Model compactness}Figure~\ref{fig:params_vs_epochs} compares the training efficiency of INN architectures and standard multilayer perceptrons (MLPs) based on the number of trainable parameters and required epochs to reach convergence. INNs constructed with 4 segments and 9 modes, 6 segments and 10 modes, 8 segments and 12 modes, and 10 segments and 14 modes achieve convergence using significantly fewer trainable parameters compared to 2-layer and 3-layer MLPs. For instance, an INN with 4 segments and 9 modes reaches convergence with fewer than 5,000 parameters and within approximately 50 epochs, whereas MLPs of comparable size require 2–3 times more parameters and substantially longer training durations. Increasing the number of segments and modes in INNs leads to improved expressive power without a proportionate increase in parameter count, highlighting the architectural efficiency of segment-mode decomposition.

\begin{figure}[h]
    \centering
    \includegraphics[width=0.9\linewidth]{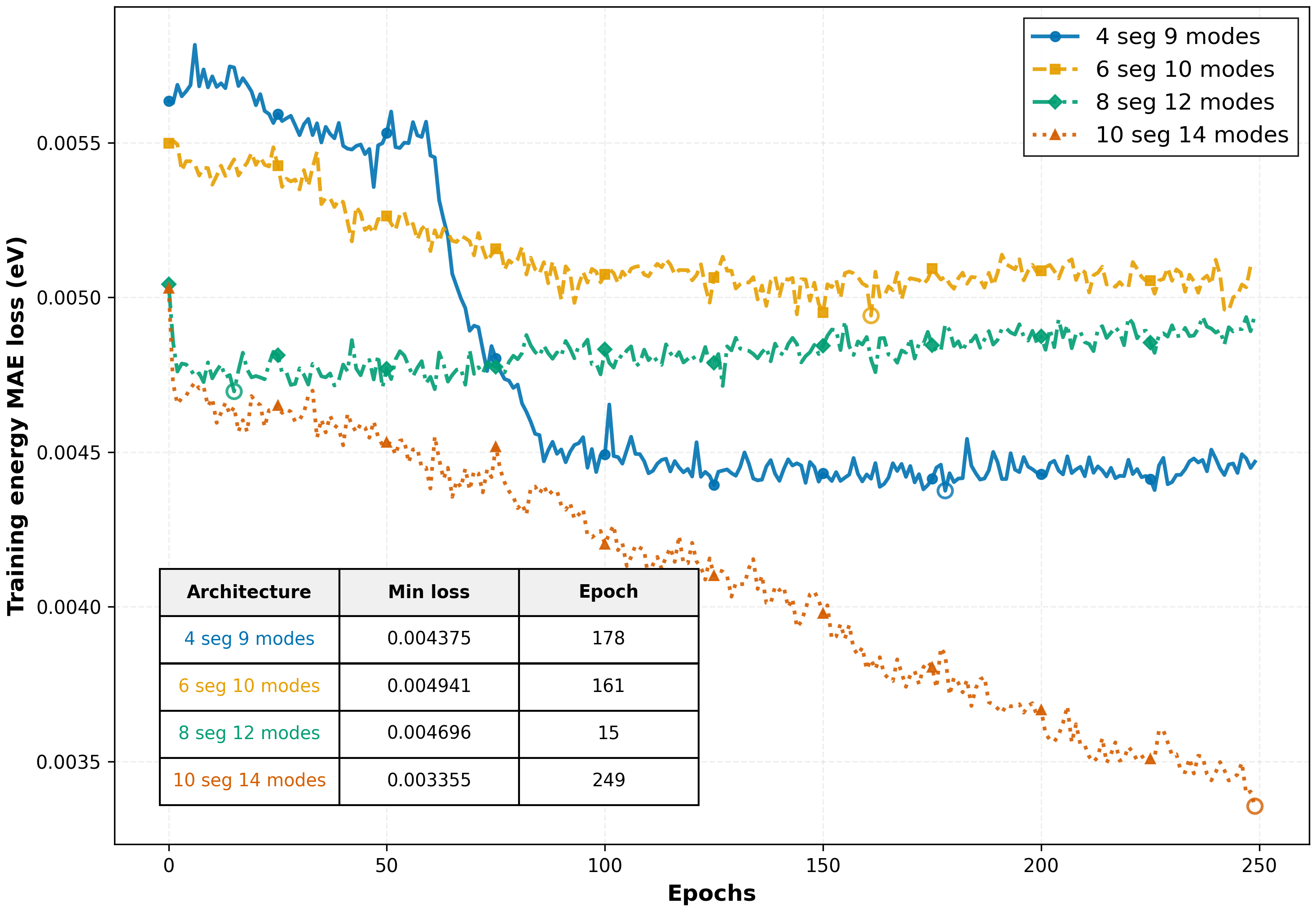}
    \caption{Training Energy Mean Absolute Error (MAE) for bulk water dataset \cite{Cheng2019}loss curves for different INN architectures over 250 epochs.}
    \label{fig:convergence_combined}
\end{figure}
\paragraph{Convergence behavior} In addition, we evaluated the impact of segment and mode configurations on the convergence behavior. Figure~\ref{fig:convergence_combined} presents the training MAE loss curves for INNs with 4 segments and 9 modes, 6 segments and 10 modes, 8 segments and 12 modes, and 10 segments and 14 modes over 250 epochs. Deeper INN architectures with more segments and modes, such as 10 segments and 14 modes, achieve the lowest final training losses, albeit requiring more epochs to converge. By contrast, configurations like 8 segments and 12 modes offer a favorable balance, reaching relatively low losses at early epochs. These results emphasize the role of segment-mode configurations in controlling the trade-off between model complexity, convergence speed, and final predictive accuracy.

\begin{figure}[h]
    \centering
    \includegraphics[width=0.7\linewidth]{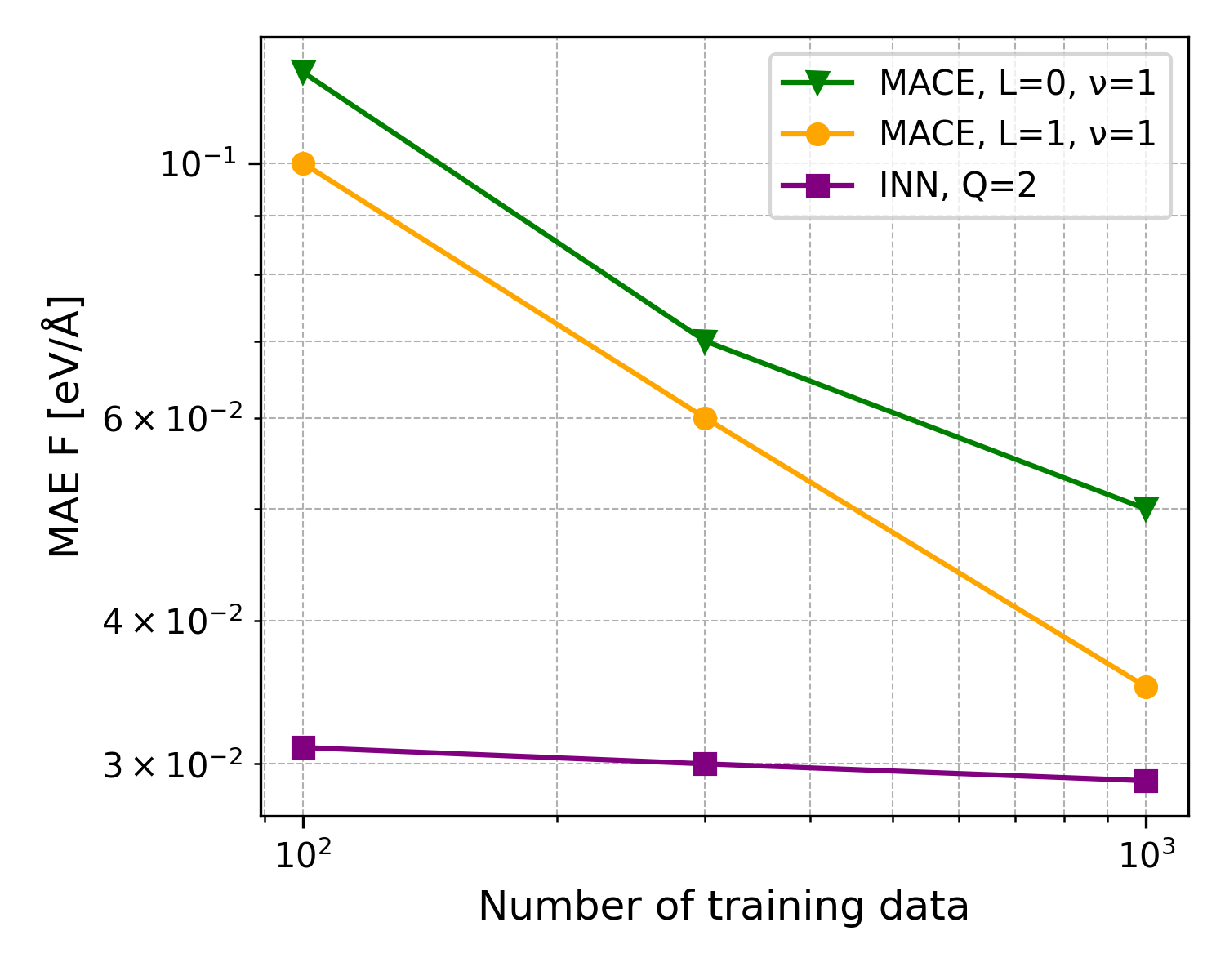}
    \caption{Learning curves for INN and MACE models trained on bulk water dataset \cite{Cheng2019}. INN demonstrates strong data efficiency across different training set sizes.}
    \label{fig:learning_curve}
\end{figure}

\paragraph{Data efficiency} Figure~\ref{fig:learning_curve} shows the force MAE learning curves for INN-FF and two MACE baselines~\cite{Batatia2023} trained on the bulk water dataset. INN-FF maintains a nearly flat curve and outperforms both MACE variants at every training set size, including the lowest (100 samples). This level of data efficiency is particularly important for ab initio datasets, where generating even a few hundred labeled configurations can take several GPU-hours. Models like INN-FF that can generalize well from limited data are more practical for atomistic simulations.

\begin{figure}[h]
    \centering
    \includegraphics[width=0.8\linewidth]{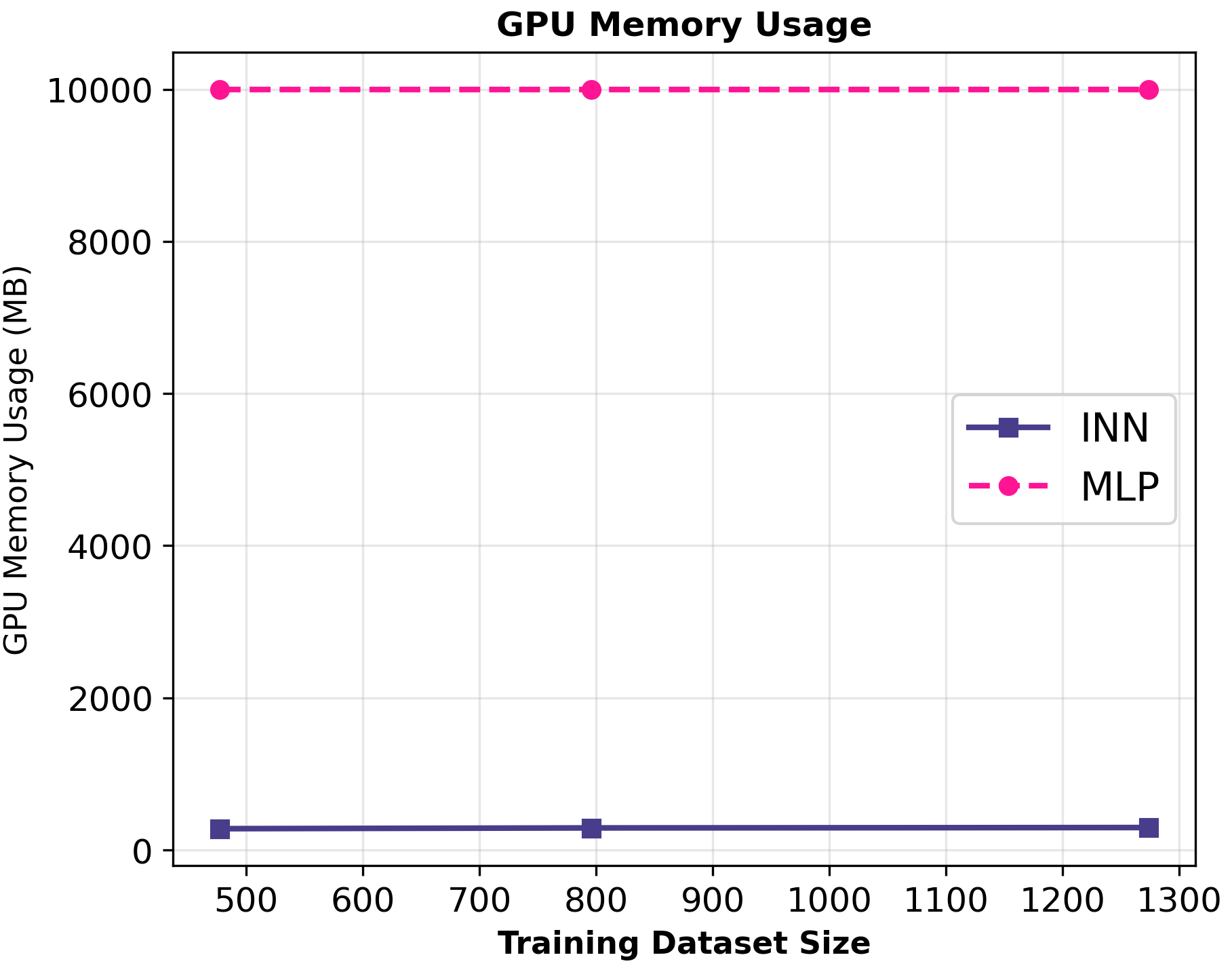}
    \caption{Comparison of training time and GPU memory usage between INN-FF and MLP models across various training dataset sizes of bulk water \cite{Cheng2019}. INN-FF exhibits consistent low memory usage and better training scalability, whereas MLPs incur high fixed memory costs that limit deployment and show diminishing performance gains.}
    \label{fig:comp_cost}
\end{figure}

\paragraph{Computational cost} Figure~\ref{fig:comp_cost} shows GPU memory usage of INN-FF and MLP models as a function of training dataset size. Across all dataset sizes, (477, 796, and 1274 samples), the MLP maintains a constant GPU memory usage of approximately 10,000MB. In contrast, INN-FF consistently uses under 300MB of GPU memory, with only a slight increase as the dataset size grows.

\subsection{Benchmark results}
\footnote{Training details for all parameters can be found in Appendix A.3.}

\subsubsection{Bulk water}
\label{benchmark}
\label{bulk_water}
We evaluated the performance of our network on a dataset consisting of 1,593 configurations of liquid water, each containing 64 molecules~\cite{Cheng2019}. The configurations were sampled from a molecular dynamics trajectory and labeled using the revPBE0-D3 level of density functional theory, implemented via the CP2K software package~\cite{Kuhne2020}. This functional is well-regarded for its ability to accurately capture the structural and dynamical behavior of water across a wide range of temperatures and pressures~\cite{Marsalek2017}.

\begin{table}[ht]
\centering
\caption{\small\textbf{Energy and force RMSE on 1,593 DFT-generated liquid water configurations with 64 molecules each} \cite{Cheng2019}}
\label{tab:bulk_water}
\vspace{0.5em}  
\scriptsize
\renewcommand{\arraystretch}{1.2}
\resizebox{\textwidth}{!}{ 
\begin{tabular}{lcccccccc}
\toprule
\textbf{Model} & \textbf{INN-FF} & \textbf{BP-NN}\cite{Cheng2019} & \textbf{REANN}\cite{zhang2021physically} & \textbf{MACE}\cite{Batatia2023} & \textbf{DEEPMD}\cite{wang2018deepmd} & \textbf{EANN}\cite{zhang2021accelerating} & \textbf{Linear ACE}\cite{witt2023acepotentials} & \textbf{CACE}\cite{cheng2024cartesian} \\
\midrule
Energy RMSE (meV/H$_2$O) & \textbf{0.133} & 7.0 & 2.4 & 1.9 & 6.3 & 6.3 & 5.196 & 3.49 \\
Force RMSE (meV/\AA)     & 42.3  & 120 & 53.2 & \textbf{37.1} & 92  & 129  & 99    & 79   \\
\bottomrule
\end{tabular}
}
\end{table}

Table \ref{tab:bulk_water} shows that INN-FF achieves the best overall accuracy, obtaining one order of magnitude lower energy RMSE among all models evaluated. The force RMSE of INN-FF is particularly notable, given the inherent difficulty of achieving low force errors in complex liquid systems such as water. MACE also demonstrates strong performance, outperforming INN slightly in force prediction but exhibiting a higher energy RMSE. CACE\cite{cheng2024cartesian} and REANN\cite{zhang2021physically} provide competitive results as well, maintaining a good balance between energy and force prediction errors. In contrast, traditional models such as BP-NN\cite{Cheng2019}, DeepMD\cite{wang2018deepmd}, and EANN\cite{zhang2021accelerating} yield substantially higher errors across both metrics.

\subsubsection{rMD17: Molecular Dynamics Trajectory}
\label{rMD17}
The revised MD17 (rMD17) dataset\cite{chmiela2017machine}, widely used for benchmarking, was trained with INN-FF for both energies and forces. It consists of 1000 random train-test splits of DFT-calculated long molecular dynamics trajectories of several small organic molecules. To highlight how effectively INN-FF trains on smaller datasets, we made the benchmark challenging by training on just 50 configurations, validating on 50 configurations, and testing on 1800 configurations. Across almost all molecules, INN-FF achieves or surpasses state-of-the-art accuracy in predicting both forces and energies.

\begin{table*}[ht]
\centering
\caption{Mean absolute errors (MAE) in energy (meV) and force (meV/\AA) on the rMD17\cite{chmiela2017machine} dataset for different molecules and methods, trained on 50 configurations}
\label{tab:rmd17_mae_final_quantity}
\vspace{0.5em}
\tiny
\renewcommand{\arraystretch}{1.3}
\resizebox{0.65\textwidth}{!}{
\begin{tabular}{llccc}
\toprule
\textbf{Molecule} & \textbf{Quantity} & \textbf{INN-FF} & \textbf{NequIP}\cite{batzner2022e3} & \textbf{MACE}\cite{Batatia2023} \\
\midrule

\multirow{2}{*}{Aspirin}
& Energy MAE (meV)     & \textbf{6.9}  & 19.5 & 17.0 \\
& Force MAE (meV/\AA)  & \textbf{20.2} & 52.0 & 43.9 \\
\midrule

\multirow{2}{*}{Ethanol}
& Energy MAE (meV)     & \textbf{4.7}  & 8.7  & 6.7  \\
& Force MAE (meV/\AA)  & \textbf{20.2} & 40.2 & 32.6 \\
\midrule

\multirow{2}{*}{Malonaldehyde}
& Energy MAE (meV)     & \textbf{4.3}  & 12.7 & 10.0 \\
& Force MAE (meV/\AA)  & \textbf{21.8} & 52.5 & 43.3 \\
\midrule

\multirow{2}{*}{Naphthalene}
& Energy MAE (meV)     & 5.1          & \textbf{2.1} & \textbf{2.1} \\
& Force MAE (meV/\AA)  & 21.7         & 10.0         & \textbf{9.2} \\
\midrule

\multirow{2}{*}{Paracetamol}
& Energy MAE (meV)     & \textbf{5.1}  & 14.3 & 9.7  \\
& Force MAE (meV/\AA)  & \textbf{21.3} & 39.7 & 31.5 \\
\midrule

\multirow{2}{*}{Salicylic acid}
& Energy MAE (meV)     & \textbf{5.9}  & 8.0  & 6.5  \\
& Force MAE (meV/\AA)  & \textbf{21.5} & 35.0 & 28.4 \\
\midrule

\multirow{2}{*}{Toluene}
& Energy MAE (meV)     & 5.4          & 3.3  & \textbf{3.1} \\
& Force MAE (meV/\AA)  & 20.9         & 15.1 & \textbf{12.1} \\
\midrule

\multirow{2}{*}{Uracil}
& Energy MAE (meV)     & 5.2          & 7.3  & \textbf{4.4} \\
& Force MAE (meV/\AA)  & \textbf{22.3} & 40.1 & 25.9 \\
\bottomrule
\end{tabular}}
\end{table*}

\section{Limitations and future work}
\label{limitations}
While INN-FF demonstrates strong performance on standard benchmarks, several limitations highlight areas for further improvement. The model has not yet been evaluated in full molecular dynamics simulations, so its long-term stability and ability to capture rare-event behavior remain untested; this can be addressed by integrating INN-FF into MD engines such as Atomic Simulation Environment\cite{larsen2017atomic} and validating its performance across extended trajectories and thermodynamic ensembles. Its generalization to more complex systems, including charged species, multi-component alloys, and reactive environments, has not been established, and may require the inclusion of additional physical descriptors or modifications to the interpolation scheme to capture long-range and many-body effects. INN-FF is also sensitive to hyperparameter choices, particularly the segment-mode configuration, and training currently relies on a narrow grid search; future versions may benefit from adaptive interpolation strategies or automated architecture tuning. Finally, although data generation was manageable for the systems studied, (primarily because the datasets were publicly available) extending INN-FF to broader chemical spaces would very likely increase computational cost. This challenge could be mitigated by incorporating active learning or multi-fidelity training strategies to reduce dependence on expensive\textit{ ab initio} data.

\section{Broader impacts}
\label{impacts}
Although it is difficult to precisely predict the societal impacts of INN-FF, its overall effect is likely to be positive, as it can contribute to accelerating drug discovery and the design of novel metal alloys. However, the computational cost associated with training the model, particularly the reliance on expensive \textit{ab initio} datasets, raises concerns regarding energy consumption and environmental impact. INN-FF addresses this challenge by generalizing well from limited high-fidelity data, thereby attempting to reduce the overall computational burden of atomistic modeling.

\section{Conclusion}
\label{conclusion}

In this work, we proposed INN-FF, a compact and scalable neural architecture for learning interatomic potentials based on interpolation and low-rank tensor representations. The model leverages locally supported shape functions and structured segment-mode decomposition to reduce parameter count and memory overhead, while maintaining high accuracy in energy and force prediction tasks. In two benchmark datasets, bulk water and revised MD17, INN-FF consistently achieves strong performance, particularly in low training data configurations. It generalizes well to unseen configurations. Compared to standard MLP baselines, INN-FF requires significantly less GPU memory and computational cost, making it well-suited for applications with limited training data or hardware constraints. Nonetheless, the current work is limited in scope. INN-FF has not yet been evaluated on more complex materials systems, such as charged or multi-component systems, and its long-timescale behavior in molecular dynamics remains unexplored. Additionally, the architecture does not explicitly enforce physical symmetries, which may affect generalization in highly symmetric domains. Despite these limitations, INN-FF offers a promising direction for building physically motivated, mesh-compatible neural potentials that balance accuracy with efficiency. We hope this work provides a foundation for future development toward robust and deployable force fields for large-scale atomistic simulations.

\newpage

\bibliographystyle{naturemag}
\bibliography{refs}


\newpage
\appendix

\section{Technical Appendices and Supplementary Material}

\subsection{INN Training Functions}

This section introduces the detailed procedure of computing compact-supported nonlinear INN interpolation functions using C-HiDeNN theory  \cite{park2023convolution}. C-HiDeNN interpolation in 1D is expressed as:

\begin{equation}
\label{eq:C-HiDeNN_1D}
\begin{aligned}
    u^h(x)=&\sum_{k\in\mathcal{N}^{\{c\}}}N^{(k)}(x)\sum_{j\in\mathcal{N}^{(k)}_s}\mathcal{W}_j^{(k)}(x)u^{(j)}=\sum_{i\in \mathcal{N}^{\{c\}}_s}{\widetilde{N}^{(i)}(x)u^{(i)}}, \quad x\in A^{\{c\}}
\end{aligned}
\end{equation}
where $A^{\{c\}}$ is the domain of the element $c$ to which the input $x$ belongs, $\mathcal{N}^{\{c\}}$ is the set of nodes in $A^{\{c\}}$, $N^{(k)}(x)$ is the linear interpolation function of node $k$ described in Figure \ref{fig:INN_1D}(a). Notice that superscripts with parentheses $()$ denote nodal indices while those with curly brackets $\{\}$ refer to elemental indices. We define a set of neighboring nodes centered at node $k$ with a patch size $s$: $\mathcal{N}_s^{(k)}$. For instance, in Figure \ref{fig:INN_1D}(b), $\mathcal{N}_{s=1}^{(k=3)}=\{x^{(2)},x^{(3)},x^{(4)} \}$, and $\mathcal{N}_{s=2}^{(k=3)}=\{x^{(1)},x^{(2)},x^{(3)},x^{(4)},x^{(5)}\}$. Similarly, let ${A}^{(k)}_s$ be a continuous domain of $\mathcal{N}_s^{(k)}$, e.g., $A^{(k=3)}_{s=2}=\{x|x^{(1)} \leq x \leq x^{(5)}\}$ in Figure \ref{fig:INN_1D}(b). Given this, $\mathcal{W}_j^{(k)}(x)$ is the convolution patch function of node $j$ among the patch nodes $\mathcal{N}^{(k)}_s$ and $u^{(j)}$ is corresponding nodal value. Finally, we can simplify the double summation into a single summation by denoting $\mathcal{N}^{\{c\}}_s=\cup_k{\mathcal{N}^{(k)}_s}$. The function $\widetilde{N}^{(i)}(x)$ is the C-HiDeNN interpolation function at node $i$.

Convolution patch functions interpolate a scalar field $u(x)$ over a domain $A^{(k)}_{s}$ using radial basis functions $R_k(x)$ and activation functions $\mathcal{A}_p(x)$:

\begin{equation}
\label{eq:convolution_patch_function}
\begin{aligned}
    u(x)=&\sum_{k\in\mathcal{N}_s^{(k)}}\mathcal{R}_{k}(x)\beta_k + \sum_{p=0}^{P}\mathcal{A}_{p}(x)\alpha_p\\
    =& \bm{\mathcal{R}}^T(x)\bm{\beta} + \bm{\mathcal{A}}^T(x) \bm{\alpha} = [\bm{\mathcal{R}}^T(x) \quad \bm{\mathcal{A}}^T(x)]\begin{bmatrix}
    \bm{\beta} \\
    \bm{\alpha} \end{bmatrix},
\end{aligned}
\end{equation}
where $p$ is the index of an activation function $\mathcal{A}_{p}(x)$. Let $K$ be $\text{n}(\mathcal{N}_s^{(k)})$. The radial basis function $\mathcal{R}_{k}(x)=\mathcal{R}(r^{(k)}(x))$ depends on the radial distance between an input coordinate $x$ and the nodal coordinate $x^{(k)}$, that is, $r^{(k)}(x)=||x-x^{(k)}||_2$. In the literature of C-HiDeNN \cite{lu2023convolution, park2023convolution}, the cubic spline function with a dilation parameter $a$ was adopted \cite{liu1995reproducing}:

\begin{equation}
\begin{aligned}
    \mathcal{R}(r^{(k)};a)=\begin{cases}
    \frac{2}{3}-4\frac{(r^{(k)})^2}{a^2} + 4\frac{(r^{(k)})^3}{a^3} & 0 \leq \frac{r^{(k)}}{a} \leq \frac{1}{2}\\
    \frac{4}{3}-4\frac{r^{(k)}}{a}+4\frac{(r^{(k)})^2}{a^2} -\frac{4}{3}\frac{(r^{(k)})^3}{a^3} & \frac{1}{2} \leq \frac{r^{(k)}}{a} \leq 1\\
    0 &\text{otherwise}
    \end{cases}
\end{aligned}
\end{equation}
The convolution patch functions can reproduce ``$P$" arbitrary activation functions $\mathcal{A}_p(x)$ where $P\leq s$. Table \ref{tab:activation} shows a list of activation functions explored in \cite{Park2024}; however, they can be other nonlinear functions in general.

\begin{table}[htbp]
  \centering
  \small 
  \caption{Choices of activation functions. $GELU(x)=x\cdot \frac{1}{2}[1+erf(\frac{x}{\sqrt{2}})]$.}
  \label{tab:activation}
  \begin{tabular}{ccccccc}
    \toprule
    & Polynomial & Sinusoidal & Exponential & Sigmoid & Tanh & GELU \\
    \midrule
    $p=0$ & $\mathcal{A}_0(x)=1$ & $=1$ & $=1$ & $=1$ & $=1$ & $=1$ \\
    \midrule
    $p=1$ & $\mathcal{A}_1(x)=x$ & $=sin(\pi x)$ & $=e^x$ & $=\frac{1}{1+e^{-x}}$ & $=\text{tanh}(x)$ & $=\text{GELU}(x)$ \\
    \midrule
    $p=2$ & $\mathcal{A}_2(x)=x^2$ & $=sin(2\pi x)$ & $=e^{2x}$ & $=\frac{1}{1+e^{-2x}}$ & $=\text{tanh}(2x)$ & $=\text{GELU}(2x)$ \\
    \midrule
    $\vdots$ & $\vdots$ & $\vdots$ & $\vdots$ & $\vdots$ & $\vdots$ & $\vdots$ \\
    \midrule
    $p=P$ & $\mathcal{A}_P(x)=x^P$ & $=sin(P\pi x)$ & $=e^{Px}$ & $=\frac{1}{1+e^{-Px}}$ & $=\text{tanh}(Px)$ & $=\text{GELU}(Px)$ \\
    \bottomrule
  \end{tabular}
\end{table}

The coefficients $\beta_k$ and $\alpha_p$ can be determined by imposing the Kronecker delta property: $u(x^{(k)})=u^{(k)}$, or, 

\begin{equation}    
\label{eq:kronecker_delta}
    \bm{u}^{K} =\bar{\bm{\mathcal{R}}}\bm{\beta} + \bar{\bm{\mathcal{A}}}\bm{\alpha},
\end{equation}
with 

\begin{equation}
    \bm{u}^{K}=\begin{bmatrix}
    u^{(1)} \\
    \vdots \\
    u^{(K)}    
    \end{bmatrix}, 
\end{equation}

\begin{equation}
    \bar{\bm{\mathcal{R}}} = \begin{bmatrix}
    \mathcal{R}(r^{(1)}(x^{(1)})) & \mathcal{R}(r^{(2)}(x^{(1)})) & \cdots & \mathcal{R}(r^{(K)}(x^{(1)})) \\
    \mathcal{R}(r^{(1)}(x^{(2)})) & \mathcal{R}(r^{(2)}(x^{(2)})) & \cdots & \mathcal{R}(r^{(K)}(x^{(2)})) \\
    \vdots & \vdots & \vdots & \vdots \\
    \mathcal{R}(r^{(1)}(x^{(K)})) & \mathcal{R}(r^{(2)}(x^{(K)})) & \cdots & \mathcal{R}(r^{(K)}(x^{(K)}))
    \end{bmatrix}, 
\end{equation}

\begin{equation}
    \bar{\bm{\mathcal{A}}} = \begin{bmatrix}
    \mathcal{A}_0(x^{(1)}) & \mathcal{A}_1(x^{(1)}) & \cdots & \mathcal{A}_P(x^{(1)}) \\
    \mathcal{A}_0(x^{(2)}) & \mathcal{A}_1(x^{(2)}) & \cdots & \mathcal{A}_P(x^{(2)}) \\
    \vdots & \vdots & \vdots & \vdots \\
    \mathcal{A}_0(x^{(K)}) & \mathcal{A}_1(x^{(K)}) & \cdots & \mathcal{A}_P(x^{(K)})
    \end{bmatrix},
\end{equation}
and the coefficients $\bm{\beta}, \bm{\alpha}$ are:

\begin{equation}
\begin{aligned}
    \bm{\beta}&=\{\beta_1,\beta_2,\cdots,\beta_K \}^T, \\
    \bm{\alpha}&=\{\alpha_0,\alpha_1,\cdots,\alpha_P \}^T .   
\end{aligned}
\end{equation}
Eq. \ref{eq:kronecker_delta} has $(K+P+1)$ unknowns with $K$ equations. To ensure a symmetric matrix equation, additional $(P+1)$ equations  $\bar{\bm{\mathcal{A}}}^T\bm{\beta}=\bm{0}$ are employed such that:

\begin{equation}
    \begin{bmatrix}
    \bm{u}^{K} \\
    \bm{0}     
    \end{bmatrix} = \begin{bmatrix}
    \bar{\bm{\mathcal{R}}} & \bar{\bm{\mathcal{A}}} \\
    \bar{\bm{\mathcal{A}}}^T & \bm{0}
    \end{bmatrix}\begin{bmatrix}
    \bm{\beta} \\
    \bm{\alpha}   
    \end{bmatrix} = 
    \bm{G}\begin{bmatrix}
    \bm{\beta} \\
    \bm{\alpha}   
    \end{bmatrix},
\end{equation}
where $\bm{G}$ being the symmetric assembled moment matrix with $(K+P+1)\times (K+P+1)$ components. Assuming it is invertible, the coefficients are computed as:

\begin{equation}
\label{eq:kronecker_coefficients}
    \begin{bmatrix}
    \bm{\beta} \\
    \bm{\alpha}   
    \end{bmatrix} = \bm{G}^{-1}\begin{bmatrix}
    \bm{u}^{K} \\
    \bm{0}     
    \end{bmatrix}.
\end{equation}
Substituting Eq. \ref{eq:convolution_patch_function} with \ref{eq:kronecker_coefficients}, it reads:

\begin{equation}
\label{eq:convolution_patch_function_w_moment}
\begin{aligned}
    u(x)&=[\bm{\mathcal{R}}^T(x) \quad \bm{\mathcal{A}}^T(x)]\bm{G}^{-1}
    \begin{bmatrix}
    \bm{u}^{K} \\
    \bm{0}     
    \end{bmatrix} \\
    &= \widetilde{\bm{\mathcal{W}}}(x)\begin{bmatrix}
    \bm{u}^{K} \\
    \bm{0}     
    \end{bmatrix}.
\end{aligned}
\end{equation}
The convolution patch functions $\bm{\mathcal{W}}(x)$ are the first $K$ components of the vector function $\widetilde{\bm{\mathcal{W}}}(x)$. 

Finally, the C-HiDeNN interpolation function in Eq. \ref{eq:C-HiDeNN_1D} becomes:

\begin{equation}
\begin{aligned}
    \widetilde{N}^{(i)}(x)&=\begin{cases}
    \sum_{k\in \mathcal{N}^{\{c\}}} N^{(k)}(x) \cdot \mathcal{W}^{(k)}_i(x) & x\in A^{\{c\}} \\
    0 &\text{otherwise},
    \end{cases}
\end{aligned}
\end{equation}

In summary, the INN interpolation functions in 1D (i.e., C-HiDeNN) are constructed with linear interpolation functions ${N}^{(j)}(x)$ and the convolution patch functions $\mathcal{W}^{(k)}_j(x)$. They satisfy the four conditions provided in Table \ref{tab:conditions}. 


\begin{table}[h!]
  \begin{center}
    \caption{\textcolor{black}{Conditions for constructing the linear interpolation functions $\mathcal{N}^{(k)}(x)$ and the convolution patch functions $\mathcal{W}^{(k)}_j(x)$.}}
    \label{tab:conditions}
    \begin{tabular}{|c|c|}
    \hline
      \textbf{Linear interpolation functions} & \textbf{Convolution patch functions} \\
      $N^{(k)}(x)$ & $\mathcal{W}^{(k)}_j(x)$ \\
      \hline
      1. Compact support  & 3. Kronecker delta:  \\ $\mathcal{N}^{\{c\}}_s=\cup_k{\mathcal{N}^{(k)}_s}$ 
        &  $\mathcal{W}^{(k)}_{j}(x^{(i)})=\delta_{ij}$ \\   \hline
      2. Partition of unity:  & 4. Reproducing conditions:\\
      $\sum_{j=1}^{J}N^{(j)}(x)=1$, $\forall x$ & $\sum_{j\in \mathcal{N}^{(k)}_s} \mathcal{W}^{(k)}_j(x) \mathcal{A}_p(x^{(j)})=\mathcal{A}_p(x)$ \\
      \hline
    \end{tabular}\\
  \end{center}
\end{table}

\subsection{Descriptions of the Datasets}

\subsubsection{Bulk Water}

 Computed using density functional theory (DFT) at the revPBE0-D3 level, the water dataset contains 1,593 configurations of liquid water. Each configuration consists of 64 molecules \cite{Cheng2019}.

\subsubsection{Revised MD17}

The original MD17 dataset was found to contain noisy labels, which led to the release of a revised version known as rMD17 \cite{chmiela2017machine}. This updated dataset covers 10 small organic molecules, with five different train-test splits for each. In every split, 1000 configurations per molecule were randomly selected from long \textit{Ab initio} molecular dynamics (AIMD) simulations at 500 K, computed using DFT.

\subsection{Training Details}

All models were trained on a single NVIDIA RTX 3060 GPU with a batch size of 1000.
For the rMD17 dataset, training was performed on 50 configurations, with an additional 50 used for validation and 1800 unseen configurations held out for testing. The model architecture employed 2 segments and a single mode. For the water dataset, 1274 configurations were used for training, 159 for validation, and 159 for testing. This model utilized 9 segments and 4 modes.

For the MLP baseline, all models were also trained on a single NVIDIA RTX 3060 GPU using a batch size of 128 and a learning rate of $10^{-2}$. The architecture consisted of three fully connected layers, each with 100 neurons, and ReLU activation functions applied after each layer. Training was conducted for 250 epochs. 

Both the INN-FF and MLP baseline frameworks were trained using JAX \cite{jax2018github}, which provided significant acceleration in training via automatic differentiation and just-in-time compilation optimized for GPU hardware.

Details of other machine-learned interatomic potentials were obtained from the literature.

\subsection{Limitations and Reproducibility}
INN-FF has not yet been benchmarked on reactive systems or those with high symmetry, and its long-term stability in molecular dynamics simulations remains to be evaluated. Additionally, full code for reproducibility cannot be shared at this time due to pending patent considerations and proprietary restrictions.

\subsection{Additional Experiments}

\begin{figure}[h]
    \centering
    \includegraphics[width=0.8\linewidth]{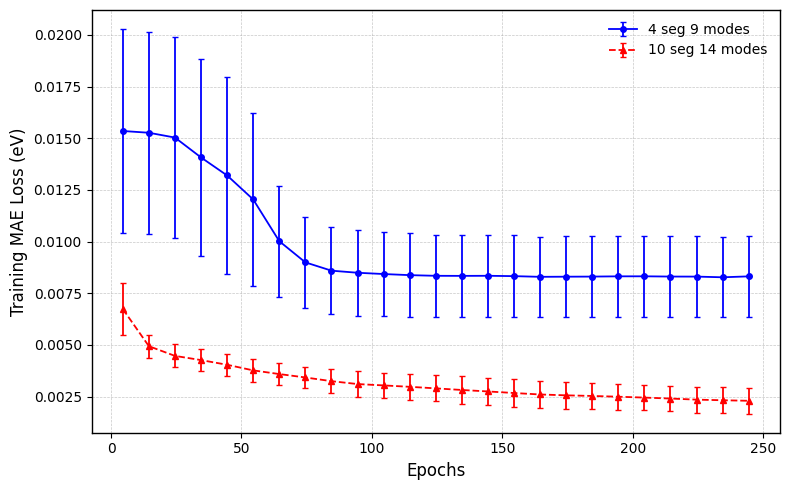}
    \caption{Training MAE loss (mean $±$ std) over 5 runs for INN-FF architectures on the bulk water\cite{Cheng2019} dataset. The 10-segment, 14-mode model (red) achieves lower final error than the 4-segment, 9-mode variant (blue), but requires more epochs to converge.}
    \label{fig:errorbar}
\end{figure}

Figure~\ref{fig:errorbar} illustrates the training convergence behavior of two INN-FF architectures on the bulk water dataset~\cite{Cheng2019}. The deeper configuration (10 segments, 14 modes) consistently achieves a lower final training mean absolute error (MAE) than the shallower 4-segment, 9-mode variant, albeit requiring more epochs to converge. Each curve represents the mean training loss over five independent training runs, where the dataset was randomly reshuffled before each run to create new training-validation splits (with consistent sample sizes). Error bars denote the standard deviation across these five runs at each 10-epoch interval.

\end{document}